\pdfoutput=1
\documentclass{sig-alternate-05-2015}
\usepackage{bm}
\usepackage{xcolor}
\usepackage{enumitem}
\usepackage{algorithm} 
\usepackage{algorithmic} 
\usepackage{caption}
\usepackage{subcaption}
\usepackage{booktabs}
\usepackage{tabularx}
\usepackage{paralist}
\defaultleftmargin{1em}{}{}{}
\usepackage[notheorems]{dgleich-math}

\usepackage{ushort}
\newcommand{\cel}[1]{\ushort{#1}}

\newcommand{\celm}[1]{\cel{\mat{#1}}}

\newcommand{\cT}{\ensuremath{\cel{T}}}

\providecommand{\cmP}{\ensuremath{\celm{P}}}

\providecommand{\cmT}{\ensuremath{\celm{T}}}
\providecommand{\cmU}{\ensuremath{\celm{U}}}

\newcommand{\cmPt}{\cel{\mat{\tilde{P}}}}
\renewcommand{\cmPt}{\mat{\tilde{P}}}

\newcommand{\elm}[1]{\cel{#1}}

\newtheorem{theorem}{Theorem}[section]

\newtheorem{observation}{Observation}

\newenvironment{definition}[1][Definition]{\begin{trivlist}
\item[\hskip \labelsep {\bfseries #1}]}{\end{trivlist}}

\newcommand{\given}

\newcolumntype{Y}{>{\raggedright\arraybackslash}X}
\newcolumntype{W}{>{\raggedleft\arraybackslash}X}
\newcolumntype{Z}{>{\centering\arraybackslash}X}

\graphicspath{{figure/}}

\begin{document}

\setcopyright{none}




 
%

\title{General Tensor Spectral Co-clustering \\ for Higher-Order Data}
%
%
%
%
%

\numberofauthors{3} 
%
\author{
%
%
\alignauthor Tao Wu \\
       \affaddr{Purdue University}\\
       \affaddr{West Lafayette, IN}\\
       \email{wu577@purdue.edu}
\alignauthor Austin R. Benson \\
       \affaddr{Stanford University}\\
       \affaddr{Stanford, CA}\\
       \email{arbenson@stanford.edu}
\alignauthor David F. Gleich\\
       \affaddr{Purdue University}\\
       \affaddr{West Lafayette, IN}\\
       \email{dgleich@purdue.edu}
}


\maketitle

\begin{abstract}
Spectral clustering and co-clustering are well-known techniques in data analysis, and recent work has extended spectral clustering to square, symmetric tensors and hypermatrices derived from a network. We develop a new tensor spectral co-clustering method that applies to any non-negative tensor of data. The result of applying our method is a simultaneous clustering of the rows, columns, and slices of a three-mode tensor, and the idea generalizes to any number of modes. The algorithm we design works by recursively bisecting the tensor into two pieces. We also design a new measure to understand the role of each cluster in the tensor. Our new algorithm and pipeline are demonstrated in both synthetic and real-world problems. On synthetic problems with a planted higher-order cluster structure, our method is the only one that can reliably identify the planted structure in all cases. On tensors based on n-gram text data, we identify stop-words and semantically independent sets; on tensors from an airline-airport multimodal network, we find worldwide and regional co-clusters of airlines and airports; and on tensors from an email network, we identify daily-spam and focused-topic sets.
\end{abstract}

%


%
%

%
%
\printccsdesc




\section{Introduction}

Clustering is a fundamental task in data mining that aims to assign closely
related entities to the same group.
Traditional methods optimize some aggregate measure of the strength
of pairwise relationships between items, such as the sum of 
similarity between all pairs of entities in the same group. 
Spectral clustering is a particularly powerful technique for computing
the clusters when the pairwise similarities are encoded into
the adjacency matrix of a graph. However, many graph-like
datasets are more naturally described by higher-order connections among
several entities.  For instance, multilayer or multiplex networks describe the
interactions between several graphs simultaneously with layer-node-node
relationships~\cite{kivela2014multilayer}.

Tensors are a common representation for many of these higher-order datasets.
A tensor, or hypermatrix, is a multidimensional array with an arbitrary
number of indices or modes. A tensor with two-modes is equivalent to a 
matrix, and a tensor with three-modes looks like a three-dimensional brick.
We recently proposed a Tensor Spectral Clustering (TSC) framework as a
generalization of spectral methods for higher-order graph
data~\cite{benson2015tensor}.  This method was designed for the case
when the higher-order tensor recorded the occurrences of small subgraph
patterns within the network. For instance, the $i,j,k$th element of the
tensor denoted the presence of a triangle or a directed $3$-cycle. The
method was particularly successful at identifying clusters that corresponded
to different layers of a directed network. 

The TSC framework had a number of limitations, however. 
First, it was primarily designed for the case when the tensor arose
based on some underlying graph. The partitioning metric used was 
designed explicitly for this case.  Thus, the applications are limited in scope and 
cannot model, for example, multiplex networks.
Second, the TSC framework involved viewing the tensor as the 
transitions on a higher-order Markov chain---similar to how spectral
clustering views pairwise data as the transitions of a first-order
Markov chain---and then using a spacey random walk on this higher-order
Markov chain to identify the clusters~\cite{Benson-preprint-srw}. 
When the data are sparse, the
spacey random walk required a correction whose magnitude is proportional
to the sparsity in the tensor. In many instances, this correction was 
substantial because the tensor was extremely sparse. This made it difficult to
accurately identify clusters.

Here we develop the General Tensor Spectral Co-clustering (GTSC) framework for
clustering general sparse tensor data.  Our method is based on a new 
\emph{super-spacey random walk} model that more accurately models 
higher-order Markov chains and avoids the correction necessary for
the previous spacey random walk model. 
Furthermore, we show how to use our method on rectangular tensor data
through a tensor symmetrization procedure~\cite{ragnarsson2013block}.  
This allows
us to simultaneously cluster the rows, columns, and slices of three-mode tensors.
The idea generalizes to any number of modes where we cluster the objects
represented in each mode independently.
We also introduce a variant on the well-known conductance measure for partitioning
graphs~\cite{schaeffer2007graph} that we call \emph{biased conductance} and describe how this provides a tensor partition quality metric.  In particular, biased conductance
is approximately the exit probability from a set following our new super-spacey random
walk model.

The algorithm underlying our GTSC framework recursively partitions the tensor
data.  We stop the process when the partitioning metric is bad, which lets
us cluster the data without specifying the number of clusters ahead of time.
This provides an advantage over existing tensor decomposition methods such as
PARAFAC that require the number of clusters as an input to the algorithms.
Finally, we show that our method asymptotically scales linearly (up to
logarithmic factors) in the size of data.  In Section~\ref{sec_complexity}, we
perform numerical scaling experiments to demonstrate the scalability of our
method.


We use extensive experiments on both synthetic and real-world problems to
validate the effectiveness of our method.  For the synthetic experiments, we
devise a ``planted cluster'' model for tensors and show that GTSC has superior
performance compared to other state-of-the-art clustering methods in recovering
the planted clusters.  In Section~\ref{sec_real_data}, we analyze tensor data
from a variety of real-world domains, including text mining, air travel, and
communication.  We find that our GTSC framework identifies stop-words and
semantically independent sets in $n$-gram tensors, worldwide and regional
airlines and airports in a flight multiplex network, and topical sets in an email
network.


Our contributions are summarized as follows:
\begin{compactitem}
\item We create a new super-spacey random walk model for tensor data and use it to create a partitioning quality measure based on a notion of biased conductance (Section~\ref{sec_high}).

\item We develop GTSC, a new framework for clustering rectangular tensor data
  (Section~\ref{sec_alg}, Algorithm~\ref{alg_1}).  The framework simultaneously clusters the rows,
  columns, and slices of the tensor.

\item We show that GTSC outperforms other state-of-the-art clustering methods
  at identifying the planted cluster structure in synthetic tensor data, in terms of
  normalized mutual information, F1 score, and adjusted rand index
  (Section~\ref{sec_synthetic_data}).

\item We empirically demonstrate that GTSC identifies coherent clusters in
  real-world datasets (Section~\ref{sec_real_data}).\footnote{\fontsize{8}{9}\selectfont Code and data for this paper are available at: \url{https://github.com/wutao27/GtensorSC}.}
\end{compactitem}



\section{FIRST-ORDER SPECTRAL METHODS}
\label{sec_first}


We first review graph clustering methods from the view of graph cuts and random
walks, and then review the standard spectral clustering method using sweep cuts.
In Section~\ref{sec_high}, we generalize these notions to higher-order
data in order to develop our GTSC framework.

\subsection{Preliminaries and Notation}


Let $\mA \in \mathbb{R}_{+}^{n \times n}$ be the adjacency matrix of an
undirected graph $G = (V, E)$ and let $n = \lvert V \rvert$ be the number of
nodes in the graph.  Define $\mD = \text{diag}(\mA\ve)$ to
be the diagonal matrix of degrees of vertices in $V$.  The graph Laplacian is
$\mL = \mD-\mA$ and the transition matrix is
$\mP = \mA^T\mD^{-1}$.  The transition matrix represents the transition
probabilities of a random walk on the graph.  If a walker is at node $j$, it
transitions to node $i$ with probability $P_{ij} = A_{ij} / D_{jj}$.

\subsection{Conductance and Markov Chains}
\label{sec_conductance}

One of the most widely-used quality metrics for partitioning a graph's vertices
into two sets $S$ and $\bar{S} = V \backslash S$ is
conductance~\cite{schaeffer2007graph}.  Intuitively, conductance measures the
ratio of the number of edges in the graph that go between $S$ and $\bar{S}$ to
the number of edges in $S$ or $\bar{S}$.  Formally, we define conductance as:
\begin{equation}
\phi(S) = \text{cut}(S)/\text{min}\big(\text{vol}(S),\text{vol}(\bar{S})\big),
\end{equation}
where
\begin{equation}
\text{cut}(S) = \sum_{\mathclap{i\in S,j\in\bar{S}}} A_{ij} \qquad \text { and } \qquad
\text{vol}(S) = \sum_{\mathclap{i\in S, j \in V}} A_{ij}.
\end{equation}
A set $S$ with small conductance is a good partition $(S, \bar{S})$.  The
numerator minimizes the edges going between the sets, and the denominator encourages either $S$ or $\bar{S}$ to be large in volume and that both are balanced.



The following well-known proposition relates conductance to random walks on the
graph.
\begin{observation}[\cite{meila2001random}]
\label{PROP}
Let $G$ be undirected, connected, and not bipartite.  Start a random walk $(X_t)_{t\in \mathbb{N}}$
where the initial state $X_0$ is randomly chosen following the stationary
distribution of the random walk.  Then for any set $S \in V$,
\[
\phi(S) = \max
\big\{
\textnormal{Pr}(X_1\in \bar{S} \mid X_0\in S),
\textnormal{Pr}(X_1\in S \mid X_0\in\bar{S})
\big\}.
\]
\end{observation}
This provides an alternative
view of conductance---it measures the probability that one step of a random walk
will traverse between $S$ and $\bar{S}$.  Again, small conductance is indicative
of a good partition: a small probability means that the sets have more internal
connections than external connections. This random walk view,
in concert with the super-spacey random walk, will serve as the basis for
our biased conductance idea to partition tensors in Section~\ref{sec_higher_order_cond}.


\subsection{Spectral Partitioning with Sweep Cuts}
\label{sec_sweep}
%

Finding the set of minimum conductance is an NP-hard combinatorial optimization
problem~\cite{wagner1993between}.  However, there are real-valued relaxations of
the problem that are tractable to solve and provide a guaranteed 
approximation~\cite{mihail1989conductance,Chung-1992-book}. The most
well known computes an eigenvector called the Fiedler vector
and then uses a sweep cut to identify a partition based on
this eigenvector. 

The Fiedler eigenvector $\vz$ solves
$\mL\vz = \lambda \mD \vz$ where $\lambda$ is the second
smallest generalized eigenvalue. 
This can be equivalently formulated in terms of the 
random walk transition matrix $\mP$. Specifically,
\[ 
\mL\vz = \lambda \mD \vz 
\,\,\,\, \Leftrightarrow \,\,\,\, (\mI - \mD^{-1} \mA) \vz = \lambda \vz 
\,\,\,\, \Leftrightarrow \,\,\,\, \vz^T \mP = (1-\lambda) \vz.
\]
In other words, the Fiedler vector is simultaneously the generalized
eigenvector with the second smallest generalized eigenvalue of the
Laplacian and degree and the left
eigenvector of $\mP$ with the second largest eigenvalue. 
This equivalence is important for our generalizations to
higher-order data in Section~\ref{sec_high}.


The \emph{sweep cut procedure} to identify a low-conductance set $S$ from $\vz$
is as follows: 
\begin{compactenum}
  \item Sort the vertices by $\vz$ as $z_{\sigma_1} \leq z_{\sigma_2} \leq \cdots \leq z_{\sigma_n}$.
  \item Consider the $n-1$ candidate sets $S_k = \{\sigma_1,\sigma_2,\cdots,\sigma_k\}$ for $1\leq k\leq n-1$
  \item Choose $S = \text{argmin}_{S_k}\phi(S_k)$ as the solution set.
\end{compactenum}
The solution set $S$ from this algorithm satisfies the celebrated Cheeger
inequality~\cite{mihail1989conductance,chung2007four}:
$\phi(S)\leq 4\sqrt{\phi_{opt}}$, where $\phi_{opt} = \min_{S \subset V} \phi(S)$
is the minimum conductance over any set of nodes.  This procedure 
is extremely efficient since $S_{k + 1}$ and $S_{k}$ differ only in the vertex
$\sigma_{k + 1}$. The conductance value of the set can be updated in time
proportional to the degree of vertex $\sigma_{k_1}$ and thus 
the sweep cut procedure only takes linear time in the number
of edges in the graph.

To summarize, the spectral method requires two components: the second left
eigenvector of $\mP$ and the conductance criterion.



\section{Higher-order spectral method}
\label{sec_high}


We now generalize the ideas from spectral graph partitioning to 
nonnegative tensor data.  We first review our
notation for tensors and then review how tensor data can be interpreted as a
higher-order Markov chain. We briefly review our prior work on Tensor Spectral
Clustering before introducing the new super-spacey random walk that we use
here. This super-spacey random walk will allow us to compute a vector akin to
the Fiedler vector for a tensor and to generalize
conductance to tensors.  Furthermore, we generalize the ideas from
co-clustering in bipartite graph data~\cite{dhillon2001co} to rectangular
tensors.

\subsection{Preliminaries and Tensor Notation}

We use $\cmT$ to denote a tensor. As a generalization of a matrix, it
may have up to $m$ indices---called an $m$-mode tensor---and 
so an individual element is 
$\cT_{i_1,i_2,\cdots,i_m}$. We will work with non-negative tensors 
where $\cT_{i_1,i_2,\cdots,i_m} \ge 0$. We call a subset of the tensor
entries with all but the first element fixed a \emph{column} of the tensor. For instance, the $j,k$ column is $\elm{T}_{:,j,k}$. A tensor is square if the dimension
of all the modes is equal and rectangular if not; a square tensor is symmetric if it is equal for any permutation of the indices. 
For simplicity in the remainder of our exposition, we will focus on the three-mode case but everything we talk about generalizes to an arbitrary number of modes. (See~\cite{gleich2015multilinear,Benson-preprint-srw} for representative examples of how the generalizations look.) We use two operations between a tensor and a vector. 
First, a tensor-vector product with a three-mode tensor can output a vector, which we denote by: 
\[ \vy = \cmT \vx^2 \quad \Leftrightarrow \quad y_i = \textstyle \sum_{j, k} \elm{T}_{i,j,k} x_j x_k. \]
Second, a tensor-vector product can also produce a matrix, which we denote by: 
\[ \mA = \cmT[\vx] \quad \Leftrightarrow \quad A_{i,j} = \textstyle  \sum_{k} \elm{T}_{i,j,k} x_k. \]

\subsection{Higher-order Markov Chains}

Recall from Section~\ref{sec_first} that we could form the transition matrix for
a Markov chain from a square non-negative matrix $\mA$ by normalizing the
columns of the matrix $\mA^T$.  We can generalize this idea to define a
higher-order Markov chain by normalizing a square tensor.  This leads to a
probability transition tensor $\cmP$:
\begin{equation}
\label{equ_tran}
\elm{P}_{i,j,k} = \elm{T}_{i,j,k} / \textstyle \sum_{i} \elm{T}_{i,j,k}
\end{equation}
where we assume $\sum_{i} \elm{T}_{i,j,k} > 0$.
(In Section~\ref{sec_sparse}, we will discuss the sparse case where the column $\elm{T}_{:,j,k}$ may not have any non-zero entries.)   

Entries of $\cmP$ can be interpreted as the transition
probabilities of a higher-order Markov chain $Z_t$:
\[
\elm{P}_{i,j,k} = \text{Pr}(Z_{t+1} = i \mid Z_t = j,Z_{t-1} = k).
\]
In terms of random walks, if the last two states were $j$ and $k$, 
then the next state is $i$ with probability 
$\elm{P}_{i,j,k}$. 


It is possible to turn any higher-order Markov chain into a first-order Markov chain on the product state space of all ordered pairs $(i,j)$. The new Markov chain moves to the state-pair $(i,j)$ from $(j,k)$ with probability $\elm{P}_{i,j,k}$. Computing the Fiedler vector associated with this chain would be one possible strategy. 
However, this approach has two immediate
problems.  First, the eigenvector is of size $n^2$, which quickly becomes
too large.  Second, the eigenvector gives information about the product
space---not the original data.  In other words, it is unclear how to use this
eigenvector even if we could afford to compute and store it.

In our prior work, we used the \emph{spacey random walk} and \emph{spacey random surfer} stochastic processes for a scalable random walk process~\cite{benson2015tensor,Benson-preprint-srw}. This stochastic process is non-Markovian and generates a sequence of states $X_t$ as follows.  After arriving at state $X_t$, the walker promptly ``spaces out'' and forgets the state $X_{t-1}$, yet it still wants to transition
according to the higher-order transitions $\cmP$. So it invents a state $Y_t$ by drawing a random state from its history and then transition to state
$X_{t+1}$ with probability $\elm{P}_{X_{t + 1}, X_{t}, Y_{t}}$.  If $H_t$ denotes the history of the process up to
time $t$,\footnote{Formally, this is the $\sigma$-algebra generated by the
  states $X_{1}, \ldots, X_{t}$.}  then
\begin{equation}\label{equ_his}
\text{Pr}(Y_{t} = j \mid H_t) = \textstyle \frac{1}{t + n}\left(1 + \sum_{r=1}^{t}\text{Ind}\{X_r = j\}\right).
\end{equation}
In this case, we assume that the process has a non-zero probability of picking any state by inflating its history count by 1 visit. 
The spacey random surfer is a generalization
where the walk follows the above process with probability $\alpha$ and
teleports at random following a stochastic vector $\vv$ with probability
$1 - \alpha$. This is akin to how the PageRank random walk includes teleportation.

These spacey random walk processes are instances of vertex reinforced random walks~\cite{Benaim-1997-vrrw,Pemantle-2007-survey}. Limiting stationary distributions are solutions to the multilinear PageRank problem~\cite{gleich2015multilinear}:
\begin{equation}
\label{equ_mp}
\alpha \cmP \vx^2 + (1-\alpha) \vv = \vx.
\end{equation} 
As shown in~\cite{Benson-preprint-srw}, the limiting distribution $\vx$
represents the stationary distribution of the transition matrix $\cmP[\vx]$.
The transition matrix $\cmP[\vx]$ asymptotically approximates the spacey walk or
spacey random surfer.  Thus, it is feasible to compute an eigenvector of this
matrix and use it with the sweep cut procedure on a generalized notion of
conductance.  However, we had to assume that all the columns of $\cmT$ were
non-zero, which does not occur in real-world datasets.  In other words, all of
the $n^2$ possible transitions from pairs of states were well defined. In our
prior work, we adjusted the tensor $\cmT$ and replaced any columns of all zeros
with the uniform distribution vector. (This is easy to do in a way that does not
require $n^2$ storage.) Because the number of zero-columns could be extremely
large, this was not ideal---although it gave promising results when the tensors
arose from graphs for a notion of conductance that was specific to these
graphs. We deal with this issue more generally in the following section,
and note that our new solution outperforms the old one as described in the
experiments. 

\subsection{A Stochastic Process for Sparse Tensors}
\label{sec_sparse}
Here we consider another model of the random surfer that avoids the issue of
undefined transitions---which correspond to columns of $\cmT$ that are all
zero---entirely. If the surfer attempts to use an undefined transition, then the
surfer moves to a random state drawn from history.  Formally, define the set of
feasible states by
\begin{equation}
\label{equ_feas_set}
\mathcal{F} = \{ (j,k) \mid \sum_{i} \elm{T}_{i,j,k} > 0 \}.
\end{equation}
Here, the set $\mathcal{F}$ denotes all the columns in $\cmT$ that are
non-zero.  The transition probabilities of our proposed stochastic process are
given by 
\begin{eqnarray}
\label{equ_spsurfer}
& & \text{Pr}(X_{t+1} = i \mid X_t=j,H_t) \\ \nonumber
& & \quad = (1-\alpha) v_i + \\ \nonumber
& & \quad \quad \alpha \text{Pr}(X_{t+1} = i \mid X_t = j, Y_t =k, H_t ) \text{Pr}(Y_t = k \mid H_t) \\
& & \text{Pr}(X_{t+1} = i \mid X_t = j, Y_t =k, H_t )  \\ \nonumber
& & \quad = \begin{cases} 
		\elm{T}_{i,j,k} / \sum_{i} \elm{T}_{i,j,k} & (j,k) \in \mathcal{F} \\
		\tfrac{1}{n+t} (1 + \sum_{r=1}^t \text{Ind}\{ X_r = i \} ) & (j,k) \not\in \mathcal{F},
	\end{cases}
\end{eqnarray}
where $v_i$ is the teleportation probability. Again $Y_t$ is chosen according to Equation~\eqref{equ_his}. We call this process the \emph{super-spacey random surfer} because when the transitions are not defined it picks a random state. 

This process is also a vertex-reinforced random walk. Let $\cmP$ be the normalized tensor $\elm{P}_{i,j,k} = T_{i,j,k} / \sum_{i} T_{i,j,k}$ only for the columns in $\mathcal{F}$ and where all other entries are zero.  
Stationary distributions of the stochastic process must satisfy the equation: 
\begin{equation} \label{equ_spmp}
\alpha \cmP \vx^2 + \alpha (1 - \normof[1]{\cmP \vx^2}) \vx + (1-\alpha) \vv = \vx,
\end{equation}  
where $\vx$ is a probability distribution vector. (See Appendix \ref{app_stationary} for the derivation.)
At least one solution must exist, which follows directly from the Brouwer fixed-point theorem. Here we give a sufficient condition for it to be unique.
\begin{theorem}
	\label{THEO_UNIQ}
	If $\alpha < 1 / (2m - 1)$  then there is a unique solution $\vx$ to \eqref{equ_spmp} for the general $m$-mode tensor.
	Furthermore, the iterative fixed point algorithm 
	\begin{equation} \label{eq:iter} \vx_{k+1} = \alpha \cmP \vx_k^2 + \alpha (1-\normof[1]{\cmP \vx_k^2}) \vx_k + (1-\alpha) \vv_k \end{equation} will converge to this solution.
\end{theorem}
(See the Appendix \ref{app_uniq} for the proof.)

%

In practice we found high values (e.g., $0.95$) of $\alpha$ did not impede convergence.
We use $\alpha = 0.8, v_i = 1/n$ for all of our experiments in this paper.


\subsection{Biased Conductance for Tensor Partitions}
\label{sec_higher_order_cond}

From Observation \ref{PROP} in Section \ref{sec_conductance}, we know that
conductance may be interpreted as the exit probability between two sets that
form a partition of the nodes in the graph.  In this section, we derive an 
equivalent first-order Markov chain from the stationary distribution of the \emph{super-spacey random
surfer}.  If this Markov chain was guaranteed to be reversible, then we could apply the standard definitions of conductance and the Fiedler vector. This will not generally be the case, and so we introduce a biased conductance measure to partition this non-reversible Markov chain with respect to starting in the stationary distribution of the super-spacey random walk. We use the second largest, real-valued eigenvector of the Markov chain as an approximate Fiedler vector. Thus, we can use the sweep cut procedure described in
Section~\ref{sec_sweep} to identify the partition. 

In the following derivation, we use the property of the two tensor-vector products: 
\[ \cmP[\vx] \vx = \cmP \vx^2. \]

The stationary distribution $\vx$ for this \emph{super-spacey random surfer} (described in Section~\ref{sec_sparse}) is equivalently the stationary distribution of the Markov chain with transition matrix 
\[ \alpha \big( \cmP[\vx] + \vx (\ve^T - \ve^T \cmP[\vx]) \big) + (1-\alpha) \vv \ve^T. \]
(To understand this expression, note that $\vx \ge 0$ and $\ve^T \vx = 1$.)
We introduce the following first-order Markov chain 
\[ \cmPt = \cmP[\vx] + \vx (\ve^T - \ve^T \cmP[\vx]). \]
This matrix represents a useful (but crude) approximation of the higher-order structure in the data. First, we determine how often we visit states using the \emph{super-spacey random surfer} to get a vector $\vx$. Then the Markov chain $\cmPt$ will tend to have a large probability of spending time in states where the higher-order information concentrates. This matrix represents a first-order Markov
chain on which we can compute an eigenvector and run a sweep cut.

We now define a biased conductance to partition it.
\begin{definition}[Biased Conductance.]
Consider a random walk $(X_t)_{t \in \mathbb{N}}$.
The biased conductance 
$\phi_{\vp}(S)$ of a set $S \subset \{1, \ldots, n\}$ is
\[
\phi_{\vp}(S) =
\max
\big\{
\text{Pr}(X_1\in \bar{S} \mid X_0\in S),\
\text{Pr}(X_1\in S \mid X_0\in\bar{S})
\big\},
\]
where $X_0$ is chosen according to a fixed distribution
$\vp$.
\end{definition}
This definition is general and has a few subtle aspects that we wish to acknowledge. First, the initial state $X_0$ is not chosen following the stationary distribution (as in the standard definition on a reversible chain) but following $\vp$ instead. This is why we call it \emph{biased conductance}. We apply this measure to $\cmPt$ using $\vp = \vx$ (the stationary distribution of the super-spacey walk). This choice emphasizes the higher-order information.

We use the eigenvector of $\cmPt$ with the second-largest real eigenvalue as an analogue of the Fiedler vector. If the chain were reversible, this would be exactly the Fiedler vector. When it is not, then it encodes largely the same type of information and is a practical heuristic. 
It is important to note that although $\cmPt$ is
a dense matrix, we can implement the two operations we need with
$\cmPt$ in time and space that depends only on the number of 
non-zeros of the sparse tensor $\cmP$ using standard
iterative methods for eigenvalues of matrices. 


\subsection{Rectangular Tensor and Co-clustering}
\begin{figure}[tbp]
\begin{center}
\includegraphics[width=0.22\textwidth]{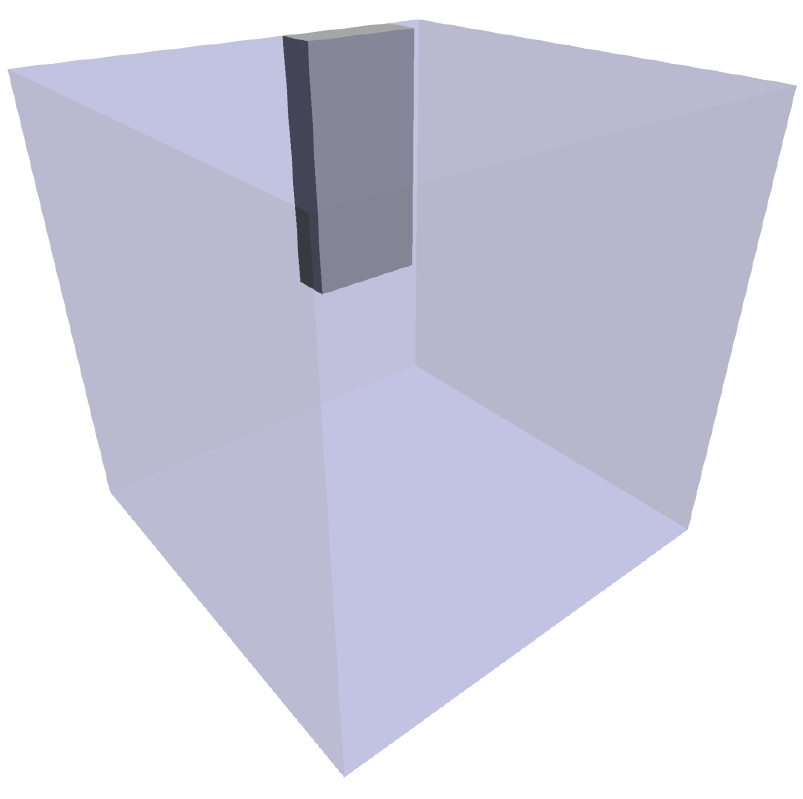}
\includegraphics[width=0.22\textwidth]{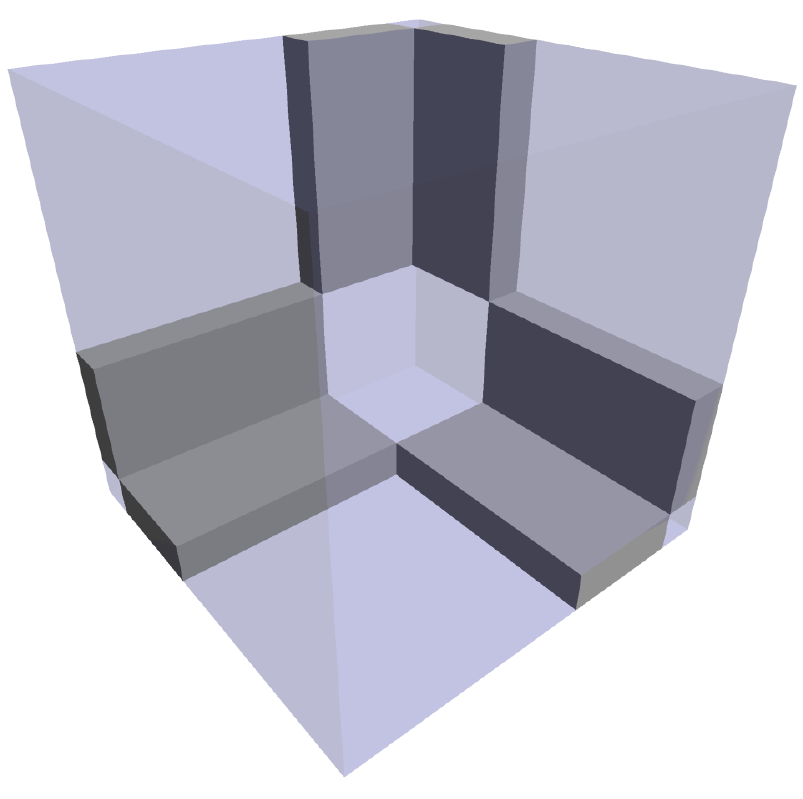}
\caption{Symmetrization of a rectangular tensor.  The tensor is first embedded into a larger square tensor (left) and then this square tensor is symmetrized (right).}
\label{fig_sym}
\end{center}
\end{figure}


So far, we have only considered square, symmetric tensor data.  However, tensor data are
often rectangular.  This is usually the case when the different modes represent
different types of data.  For example, in Section~\ref{sec_real_data}, we
examine a tensor $\cmT \in \mathbb{R}^{p \times n \times n}$
of airline flight data, where $\elm{T}_{i,j,k}$ represents that there
is a flight from airport $j$ to airport $k$ on airline $i$.  Our approach is
to embed the rectangular tensor into a larger square tensor and then symmetrize
this tensor, using approaches developed by Ragnarsson and Van
Loan~\cite{ragnarsson2013block}.  After the embedding, we can run our algorithm
to simultaneously cluster rows, columns, and slices of the tensor.  This
approach is similar in style to the symmetrization of bipartite graphs for
co-clustering proposed by Dhillon~\cite{dhillon2001co}.


Let $\cmU$ be an $n$-by-$m$-by-$\ell$ rectangular tensor. Then we embed $\cmU$ into a square three-mode tensor $\cmT$ with $n+m+\ell$ dimensions and where 
$\elm{U}_{i,j,k} = \elm{T}_{i,j+n,k+n+m}$. This is illustrated in Figure~\ref{fig_sym} (left). Then we symmetrize the tensor by using all permutations of the indices Figure~\ref{fig_sym} (right). 
The result is, when viewed as a 3-by-3-by-3 block tensor,
\[ 
\cmT = 
\left[ \begin{array}{c|c|c} 
\begin{smallmatrix}  0 & 0 & 0 \\ 
0 & 0 & \cmU_{(2,3,1)} \\
0 & \cmU_{(3,2,1)} & 0 
\end{smallmatrix} 
& 
\begin{smallmatrix}  0 & 0 & \cmU_{(1,3,2)} \\ 
0 & 0 & 0 \\
\cmU_{(3,1,2)} & 0 & 0  \\
\end{smallmatrix} 
& 
\begin{smallmatrix}  0 & \cmU_{(1,2,3)} & 0 \\ 
\cmU_{(2,1,3)} & 0 & 0 \\
0 & 0 & 0  \\
\end{smallmatrix} 
\end{array} 
\right].
\]
The tensor $\cmU_{(1,3,2)}$ is just a generalized transpose of $\cmU$ with the dimensions permuted.

Finally, we note that ``rectangular'' tensors are not determined by sizes of
modes only.  We may consider a tensor $\cmT$ to be rectangular if each mode
represents a different type of object.  The important idea is that if a tensor
is declared to be rectangular, then the result of clustering is a subset of each mode.


\section{THE ALGORITHM}
\label{sec_alg}

In this section, we put together the pieces from Section~\ref{sec_high} to build
the GTSC framework and analyze its computational complexity.  In
Section~\ref{sec_rank_clusters} we also derive a ``popularity'' metric for clusters
that will be useful for discussing clustering results on real-world data in
Section~\ref{sec_real_data}.

\subsection{Recursive Two-way Cuts}

%
%
%
%
%
%
%
%
%
%
%

\begin{algorithm}[tbp]
\caption{General Tensor Spectral Co-clustering}
\label{alg_1}
\begin{algorithmic}[1]
\REQUIRE ~~\\
Symmetric square tensor $\cmT\in \mathbb{R}_+^{n \times n \times n}$, $\alpha \in (0,1)$\\
Stopping criterion $\texttt{max-size}, \texttt{min-size}, \phi^*$
\ENSURE ~~\\
Partitioning $C$ of indices $\{1, \ldots, n\}$. \\

\STATE $C = \{ \{1, \ldots, n\} \}$
\STATE IF $n\leq \texttt{min-size}$:
\quad \textbf{RETURN}

\STATE Generate transition tensor $\cmP$ by
\begin{align}
\elm{P}_{ijk}
&=%
\begin{cases}
\elm{T}_{ijk} / \sum_{i = 1}^{n}\elm{T}_{ijk} & \text{if } \sum_{i = 1}^{n}\elm{T}_{ijk} > 0 \\
0 & \text{otherwise}
\end{cases} \nonumber
\end{align}

\STATE Compute super-spacey stationary vector $\vx$ (Equation \eqref{equ_spmp})
and form $\cmP[\vx]$.

\STATE Compute second largest left, real-valued eigenvector $\vz$ of \\$\cmPt = \cmP[\vx] + \vx(\ve^T  - \ve^T \cmP[\vx])$ (that is, $\vz^T \cmPt = \lambda \vz^T$).

\STATE $\sigma\leftarrow$ Sort eigenvector $\vz$

\STATE $(S,\phi_{\vp}) \leftarrow \text{Biased Conductance Sweep Cut} (\sigma, \cmP[\vx])$ with bias $\vp = \vx$.

\IF{$n\geq \texttt{max-size}$ or $\phi_{\vp} \leq \phi^*$}
\STATE $C_S =$ Algorithm \ref{alg_1} on sub-tensor $\cmT_{S, S, S}$. \\
\STATE $C_{\bar{S}} =$ Algorithm \ref{alg_1} on sub-tensor $\cmT_{\bar{S}, \bar{S}, \bar{S}}$. \\
\STATE $C = C_S \cup C_{\bar{S}}$.
\ENDIF

\STATE \textbf{RETURN $C$}

\end{algorithmic}

\end{algorithm}


Our GTSC algorithm works by recursively applying the sweep cut
procedure, similar to the recursive bisection procedures for clustering
matrix-based data~\cite{boley1998principal}.  We continue partitioning as long
as the clusters are large enough or we can get good enough splits.
Specifically, if a cluster has dimension less than a specified size
$\texttt{min-size}$, we do not consider it for splitting.  Otherwise, the
algorithm recursively splits the cluster if either (1) its dimension is above
some threshold $\texttt{max-size}$ or (2) the biased conductance of a new
split is less than a target value $\phi_*$.  The overall algorithm is summarized
in Algorithm~\ref{alg_1}.

We also have a couple of pre-processing steps.  First, we have to symmetrize the
data if the tensor is rectangular.  Second, we look for ``empty" indices that do
not participate in the tensor structure.  Formally, index $i$ is empty if
$\cmT_{ijk} = 0$ for all $j$ and $k$.

\subsection{Computational Complexity}
\label{sec_complexity}



We now provide an analysis of the running time of our algorithm.  Let $N$ be the
number of non-zeros in the tensor $\cmT$.  First, note that the pre-processing
(tensor symmetrization and finding empty nodes) takes $O(N)$ time.  Now, we
examine the computational complexity of a single partition:

\begin{compactenum}
  \item Generating the transition tensor $\cmP$ costs $O(N)$.
  \item Each step of \eqref{eq:iter} to find the stationary distribution is $O(N)$.
  \item Constructing $\cmP[\vx]$ costs $O(N)$. (The matrix
  $\tilde{\cmP}$ is not formed explicitly).
  \item Each iteration of the eigenvector computation takes time
  linear in the number of non-zeros in $\cmP[\vx]$, which is $O(N)$.
  \item Sorting the eigenvector takes $O(n \log n)$ computations, which is negligible considering $N$ is big compared to $n$.
  \item The sweep cut takes time $O(n + N)$, which is $O(N)$.
\end{compactenum}

In practice, we find that only a few iterations are needed to compute the
stationary distribution, which is consistent with past
results~\cite{benson2015tensor,gleich2015multilinear}.  For these systems, we do
not know how many iterations are needed for the eigenvector computations.
However, for the datasets we analyze in this paper, the eigenvector computation
is not prohibitive.  Thus, we can think of the time for each cut as roughly
linear in the number of non-zeros.  Provided that the cuts are roughly balanced,
the depth of the recursion tree is $O(\log N)$, and the total time is $O(N\log
N)$.  Again, in our experiments, this is the case.



\begin{figure}[tb]
\begin{center}
  \includegraphics[height=1.5in]{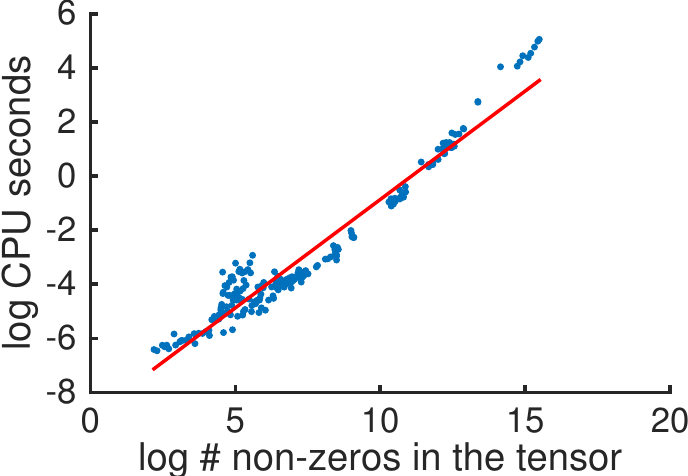}
  \caption{Time to compute a partition on the English $3$-grams as a function
  of the number of non-zeros in the tensors.  We see
  that the algorithm scales roughly linearly in the number of non-zeros (the red line).}
  \label{fig_scal}
\end{center}
\end{figure}

To backup our analysis, we tested the scalability of our algorithm on a data
tensor of English $3$-grams (see
Section~\ref{sec_real_data} for a full description of the dataset).  We varied
the number of non-zeros in the data tensor from five million down to a few
hundred by removing non-zeroes uniformly at random.  We used a laptop with $8$GB of memory and $1.3$GHz of CPU to run Algorithm~\ref{alg_1} on these 
tensors with $\texttt{max-size} = 100$,
$\texttt{min-size} = 5$, and $\phi^* = 0.4$.  Figure~\ref{fig_scal} shows the
results, and we see that the scaling is roughly linear.

\subsection{Ranking Clusters with Popularity Scores}
\label{sec_rank_clusters}

%

For our analysis of real-world datasets in Section~\ref{sec_real_data}, it will
be useful to order clusters based on the volume of their interactions with other
clusters.  To this end, we compute a ``popularity score'' for each cluster.  Let
$k$ be the number of clusters and define a $k \times k$ interaction matrix $\mM$
by the total tensor weight between the clusters.  Formally, for a three-mode
tensor, $M_{ij} = \sum_{i \in S_i, j \in S_j, k} \elm{T}_{ijk}$.  Finally,
define the popularity score of cluster $i$ as the PageRank score of node $i$ in
the graph induced by $\mM$.  (To handle corner cases, isolated nodes in this
graph are given a popularity score of 0).  For our experiments, we compute the
PageRank score with $\alpha = 0.99$ and uniform teleportation vector.


\section{Synthetic experiments}
\label{sec_synthetic_data}


We generate tensors with planted cluster structures and try
to recover the planted clusters.  We compare our GTSC framework with a variety
of other methods and find that it recovers the planted structure most often.
\subsection{Synthetic Dataset Generation}

\begin{table*}[tb]
\caption{Results of various clustering methods on the synthetically generated data.
Bold indicates the best mean performance in terms of ARI, NMI, or F1 score, and $\pm$ entries
are the standard deviations over 5 trials.  Our
method (GTSC) has the best performance in all cases.}
\vspace{-0.5cm}
\label{table_syn}
\begin{center}
\begin{tabularx}{\linewidth}{>{\hsize=1.4\hsize}Z Z Z Z Z Z Z}
\toprule
 & ARI & NMI & F1 & ARI & NMI & F1 \\
\cmidrule{2-7}
 &\multicolumn{3} {>{\hsize=3\hsize}Z}{Square tensor with $\sigma = 4$} &\multicolumn{3} {>{\hsize=3\hsize}Z}{Rectangular tensor with $\sigma = 4$} \\
\midrule
GTSC  & $\bm{0.99} \pm 0.01$ &$\bm{0.99} \pm  0.00$ & $\bm{0.99} \pm 0.01$ & $\bm{0.97} \pm 0.06$ & $\bm{0.98} \pm 0.03$ & $\bm{0.97} \pm 0.05$\\
TSC & $0.42\pm 0.05$ & $0.60\pm 0.04$ &$0.45\pm 0.04$ & $0.38\pm 0.17$ & $0.53\pm 0.15$ & $0.41\pm 0.16$\\
PARAFAC & $0.82 \pm 0.05$ & $0.94 \pm 0.02$ &$0.83 \pm 0.04$ & $0.81 \pm 0.04$ & $0.90 \pm 0.02$ & $0.82 \pm 0.04$\\
Spectral Clustering & $\bm{0.99} \pm 0.01$ & $\bm{0.99} \pm  0.01$ &$\bm{0.99} \pm  0.01$ & $0.91 \pm 0.06$ & $0.94 \pm 0.04$ & $0.91 \pm 0.06$\\
\midrule
&\multicolumn{3} {>{\hsize=3\hsize}Z}{Square tensor with $\sigma = 2$} &\multicolumn{3} {>{\hsize=3\hsize}Z}{Rectangular tensor with $\sigma = 2$} \\
\midrule
GTSC  & $\bm{0.78} \pm 0.13$ &$\bm{0.89} \pm 0.06$ & $\bm{0.79} \pm 0.12$ & $\bm{0.96} \pm 0.06$ & $\bm{0.97} \pm 0.04$ & $\bm{0.96} \pm 0.06$\\
TSC & $0.41\pm 0.11$ & $0.60\pm 0.09$ &$0.44\pm 0.10$ & $0.28\pm 0.08$ & $0.44\pm 0.10$ & $0.32\pm 0.08$\\
PARAFAC & $0.48 \pm 0.08$ & $0.67 \pm 0.04$ &$0.50 \pm 0.07$ & $0.10 \pm 0.04$ & $0.24 \pm 0.05$ & $0.15 \pm 0.04$\\
Spectral Clustering & $0.43 \pm 0.07$ & $0.66 \pm 0.04$ &$0.47 \pm 0.06$ & $0.38 \pm 0.07$ & $0.52 \pm 0.05$ & $0.41 \pm 0.07$\\
\bottomrule
\end{tabularx}
\end{center}
\end{table*}



\textbf{Square Tensor Data.}
%
%
We first generate 20 groups of nodes that will serve as our planted clusters.
The number of nodes in each group $n_g$ is drawn from a normal distribution with mean
20 and variance 5, with
truncated minimum value so that each group has at least 4 nodes.  Our square
synthetic data tensor will have mode $3$ and dimension $\sum_{g=1}^{20}n_g$.
For each group $g$ we
also assign a weight $w_g$ where the weight depends on the group number.
For group $i$, the weight is $(\sigma \sqrt{2\pi})^{-1} \exp(-(i-10.5)^2/(2\sigma^2))$, 
where $\sigma$ varies by experiment.
With these weights, groups $10$ and $11$ have the largest weight and groups $1$
and $20$ have the smallest weight. 
As described next, these weights will be
used to provide a skew in the distribution of the number of interactions for a
given dimension.


Non-zeros correspond to interactions between three indices, and we
call these assignments triples.  We generate $t_{w}$ triples whose indices
are \emph{within} a group and $t_{a}$ triples whose indices span \emph{across}
more than one group.  The $t_{w}$ triples are chosen by first uniformly
selecting a group $g$ and then uniformly selecting three indices $i$, $j$, and
$k$ from group $g$ and finally assigning a weight of $w_g$.  Formally, the value
of the tensor data tensor $\cmT$ is
\[
\elm{T}_{i,j,k} = w_g.
\]
If the same three indices are chosen more than once in the sampling procedure,
we simply increment the value in the tensor.  For the $t_{a}$ triples that
span multiple groups, the sampling procedure first selects an index $i$ from
group $g_i$ proportional to the weights of the group.  In other words, indices
in group $g$ are chosen proportional to $w_g$.  Two indices $j$ and $k$ are then
selected uniformly at random from groups $g_j$ and $g_k$ other than $g_i$.
Finally, the weight in the tensor is assigned to be the average of the three
group weights:
\[
\elm{T}_{i,j,k} = (w_{g_i} + w_{g_j} + w_{g_{k}}) / 3.
\]

For our experiments, $t_{w} = 10,000$ and $t_{a} = 1,000$, and the variance
$\sigma$ that controls the group weights is $2$ or $4$.
For each value of $\sigma$, we create $5$ sample datasets.

\textbf{Rectangular Tensor Data.}
%
For rectangular data, we distinguish between the indices for each mode of our
mode-$3$ tensor.  We label the modes as $x$, $y$, and $z$.  The groups are
generated by mode subgroups of size $n^x_g$, $n^y_g$, and $n^z_g$, $1 \le g \le
20$.  Each subgroup size is sampled from the same normal distribution as before
 with the same truncated minimum
value.  The dimension of the data
tensor is then $\sum_{g=1}^{20}n^x_g \times \sum_{g=1}^{20}n^y_g \times
\sum_{g=1}^{20}n^z_g$.  Each group $g$ is assigned a weight in the same way as
the synthetic square data.

Index triples $(i, j, k)$ now correspond to three subgroups $g^x_i$,
$g^y_j$, and $g^z_k$, and we generate within-group and across-group triples
similarly to the square case.  The $t_{w}$ within-group triples are chosen by
first uniformly selecting a group and then uniformly selecting one index from
each subgroup. The $t_{a}$ across-group triples are chosen by (1) selecting a
mode ($x$, $y$, or $z$) uniformly at random, (2) selecting the mode index
proportional to the weights $w_g$, and (3) selecting the other two indices
uniformly at random from the other groups. The value of triple $t_{w}$ and $t_{a}$
is assigned in a similar manner to the square tensor case.

We again set $t_{w} = 10,000$ and $t_{a} = 3,000$ and generate $5$ synthetic
datasets for each value of $\sigma \in \{2, 4\}$.

\subsection{Clustering Methods and Evaluation}

We compared the results of our GTSC framework to several other state-of-the-art
methods for clustering tensor data. 

\textbf{GTSC.} This is the method presented in this paper
(Algorithm~\ref{alg_1}).  We use the parameters $\texttt{max-size} = 100$, $\texttt{min-size} = 5$, and
$\phi^* = 0.35$.\footnote{ We tested several values $\phi^* \in [0.3, 0.4]$ and
  obtained roughly the same results.}

\textbf{TSC.} This is the original tensor spectral clustering
algorithm~\cite{benson2015tensor}.  We use the algorithm with recursive
bisection to find $20$ clusters.


\textbf{PARAFAC.} The PARAFAC method is a widely used tensor
decomposition procedure~\cite{harshman1970foundations} that finds an
approximation to the tensor by the sum of outer-products of vectors.  We compute
a rank-$20$ decomposition using the Tensor Toolbox \cite{TTB_Software,chi2012tensors}, and
then assign nodes to clusters by taking the index of the vector with highest
value in the nodes index.  We use the default tolerance of $10^{-4}$ and a
maximum of $1000$ iterations.


\textbf{Spectral Clustering.}  Our clustering framework (Algorithm~\ref{alg_1})
works on mode-$2$ tensors, i.e., matrices.  In this case, with $\alpha = 1$, our algorithm reduces
to a standard spectral clustering method.  We create a matrix $\mM$ from the
tensor data $\cmT$ by summing along the third mode:
$M_{ij} = \sum_{k=1}^n \elm{T}_{i,j,k}$.  We then run Algorithm~\ref{alg_1} with the same
parameters as GTSC.

\textbf{Evaluation metrics.}  We evaluate the clustering results using the
Adjusted Rand Index (ARI)~\cite{hubert1985comparing}, Normalized Mutual
Information (NMI)~\cite{lancichinetti2009detecting}, and F1 score.  The ground
truth labels correspond to the generated groups.

\subsection{Experimental Results}

Table~\ref{table_syn} depicts the performances of the four algorithms.  In all
cases, GTSC has the best performance.  In the square tensor case when $\sigma = 4$,
the standard spectral method performs as well as GTSC.  However, when
$\sigma = 2$, the score drops for all four methods with the least impact on
GTSC and TSC.  Both PARAFAC and standard spectral clustering lose nearly half of
their performances. Similar conclusions hold for the rectangular case.


The value of $\sigma$ affects the concentration of the weights and how certain
groups of nodes interact with others.  When $\sigma$ is small, weights in the
middle groups (e.g., $10$ and $11$) are very high, and many across-group
triangles have an index in these groups.  This skew reflects properties of the
real-world networks we examine in the next section.  For example, with $n$-grams
in text data, stop words such as `a', `the', and `we' are responsible for the
majority of connections in the tensor.  As evidenced by the results in
Table~\ref{table_syn}, our GTSC framework handles this skew much better than
other methods.


\section{Real-world Experiments}
\label{sec_real_data}

We now use our GTSC framework to cluster real-world tensor datasets.
Table~\ref{table_data} describes the relevant statistics of these datasets.

\begin{table}[htbp]
\caption{Statistics of tensor datasets.}
\vspace{-0.5cm}
\label{table_data}
\begin{center}
\begin{tabularx}{\linewidth}{cZc}
\toprule
Dataset & Size & \# non-zeros \\
\midrule
Airline-airport & $539\times 2,939\times2,939$ & $51,982$ \\
English $3$-grams & square 30,966 & 1,020,009 \\
Chinese $3$-grams & square 18,387 & 966,138 \\
English $4$-grams & square 23,734 & 1,034,307 \\
Chinese $4$-grams & square 14,728 & 1,002,660 \\
Enron email & $185\times 184\times 184\times 34$ & $61,891$ \\
\bottomrule
\end{tabularx}
\end{center}
\end{table}


\textbf{Airline-airport.}  This dataset consists of global air flight routes
from $539$ airlines and $2,939$ airports from
OpenFlights\footnote{http://openflights.org/data.html\#route}.  The $539 \times
2,939 \times 2,939$ data tensor $\cmT$ summarizes flights connecting airports
on several airlines.  Formally, $\elm{T}_{ijk}$ is 1 if airline $i$ flies between
airports $j$ and $k$ and $0$ otherwise.

\textbf{English 3,4-grams.}  An $n$-gram is a contiguous sequence of $n$ words
from a sequence of text.  We generated a tensor dataset from the one million most frequent
$3$-grams and $4$-grams from the Corpus of Contemporary American
English.\footnote{\url{http://www.ngrams.info/intro.asp}}\footnote{The datasets
  contain ties in frequency, so the actual number of $n$-grams is only roughly 1
  million---see Table~\ref{table_data}.} The $3$-gram tensor entry
$\elm{T}_{ijk}$ is given by the number of times that words $i$, $j$, and $k$
co-occur in an $n$-gram in the corpus.  The $4$-gram tensor dataset is defined
analogously. We symmetrize these tensors for our method.

\textbf{Chinese 3,4-grams.}  We also constructed $n$-gram tensor datasets from
the Chinese language using the Google Books $n$-grams
dataset.\footnote{https://books.google.com/ngrams} We constructed $3$- and
$4$-gram tensors in the same way as the English dataset.


\textbf{Enron email.}  This dataset is constructed from emails between Enron
employees with labeled topics~\cite{berry2001topic}.  The tensor data represents
the volume of communication between two employees discussing a given topic
during a particular week.  In total, there are $185$ weeks of data, $184$
employees, and $34$ topics, leading to a $185 \times 184 \times 184 \times 34$
tensor where
$\elm{T}_{ijkl}$ is the number of emails between employee $j$ and $k$ on topic $l$ during week $i$.


In all of our experiments, we use the stopping criterion $\phi^* = 0.4$ for
Algorithm~\ref{alg_1}.  For the $n$-gram data and airline-airport data, we use
the parameters $\texttt{max-size} = 100$ and $\texttt{min-size} = 5$; for the Enron email data,
we use $\texttt{max-size} = 50$ and $\texttt{min-size} = 10$.

\subsection{Airline-airport}
\begin{figure}
	\begin{center}
		\includegraphics[width=0.5\linewidth]{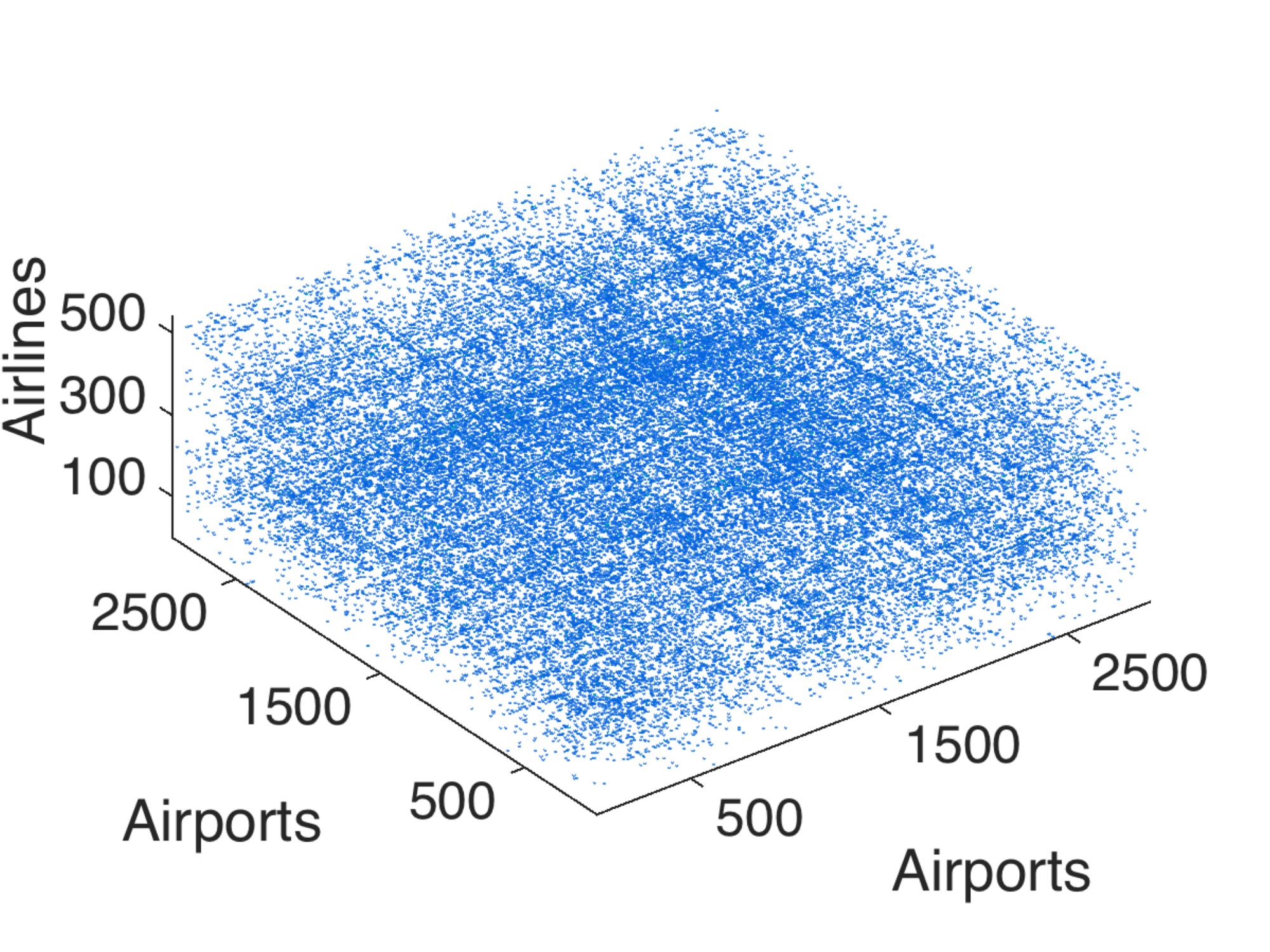}%
		\includegraphics[width=0.5\linewidth]{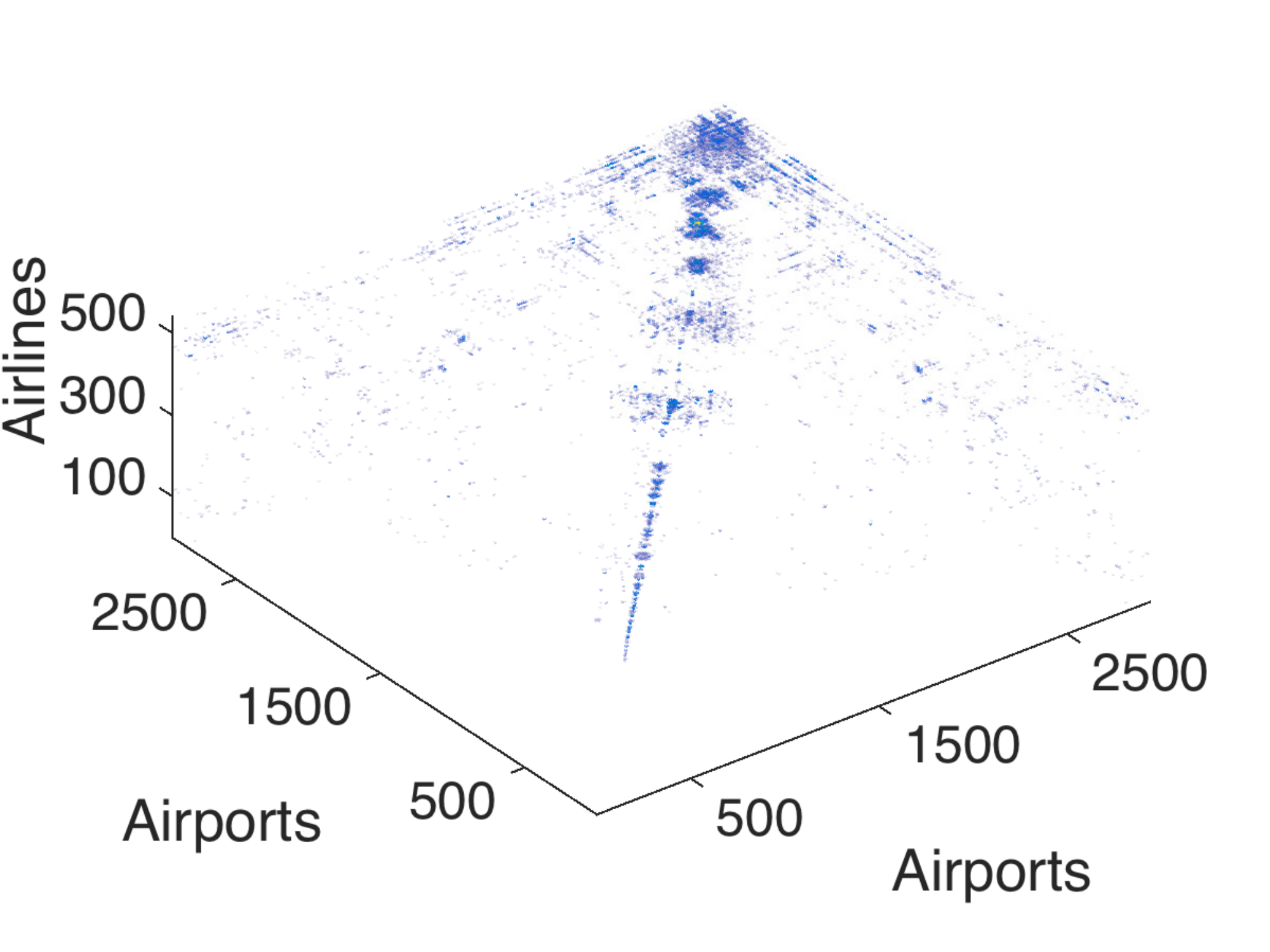}
		\vspace{-0.15cm}
		\caption{Visualization of the Airline-airport data tensor.  The $x$ and $y$ axes index airports
			and the $z$ axis indexes airlines.  A dot represents that an airline flies between those
			two airports.  On the left, indices are sorted randomly.  On the right, indices are sorted
			by the co-clusters found by our GTSC framework, which reveals structure in the tensor.}
		
		\label{fig_flight}
	\end{center}
\end{figure}


Our GTSC framework groups the airports and airlines into $129$ co-clusters.
Figure~\ref{fig_flight} illustrates the connectivity of the tensor with a random
ordering of the indices (left) and the ordering given by the popularity of co-clusters (right).
We can see that after the co-clustering, there is clear structure in the
data tensor.  Table~\ref{table_flights} summarizes some of the larger clusters found
by our method, and we discuss these clusters in more detail below.

\begin{table*}[tb]
	\caption{High-level descriptions of the larger co-clusters found by our
		GTSC framework on the Airline-airport dataset.  The algorithm finds one
		co-cluster of international hubs and large commercial airlines and
		several geographically coherent groups.}
	\label{table_flights}
	\vspace{-0.5cm}
	\begin{center}
		\begin{tabularx}{\linewidth}{>{\hsize=1.14\hsize}X >{\hsize=0.5\hsize}Z >{\hsize=0.47\hsize}Z >{\hsize=1.5\hsize}X >{\hsize=1.4\hsize}X}
			\toprule
			Name &  \# Airports & \# Airlines & Airports description & Airlines description \\
			\midrule
			Worldwide metropolises & 250 & 77 & 
			Large hubs, e.g., Beijing Capital and JFK in New York&
			Large commercial airlines, e.g., United, Air Canada, Air China \\
			Europe & 184 & 32 &
			177 in Europe, rest in Morocco & 29 European airlines \\
			United States & 137 & 9 &
			136 in U.S., Canc\'{u}n International & 29 all U.S. airlines \\
			China/Taiwan & 170 & 33 &
			136 in China or Taiwan,  & 21 in China/Taiwan 14 in S.\ Korea and Thailand\\
			Oceania/S.E.\ Asia & 302 & 52 &
			231 in Oceania or S.E.\ Asia, & 41 in East Asia or Canada 66 in China, Japan, or Canada\\
			Mexico/Americas & 399 & 68 &
			396 in Mexico or Central and South America&
			43 in Mexico or Central and South America \\
			\bottomrule
		\end{tabularx}
	\end{center}
\end{table*}


The co-cluster with the highest popularity score
is the one marked as
\textit{Worldwide Metropolises}.  This group is composed of large international
airports in cities such as Beijing and New York City.  The 250 airports in this
group are responsible for 59\% of the total routes, even though they only
account for 8.5\% of the total number of airports.  Figure~\ref{fig_flight}
illustrates this result---the airports with the highest indices are connected to
almost every other airport.  This cluster is analogous to the ``stop word'' group
we will see in the $n$-gram experiments.

%


Groups with medium levels of popularity are organized geographically.  Our GTSC
framework found one large cluster for Europe, the United States, China/Taiwan,
Oceania/SouthEast Asia, and Mexico/Americas.  Interestingly, Canc\'{u}n
International Airport is included with the United States cluster, probably due
to the large amounts of tourism.  In addition to the large groups, we find many
mini groups which consist of 5--30 airports.  These airports compose the long,
concentrated diagonal in Figure~\ref{fig_flight}.  The airports and airlines in
these co-clusters are also closely connected geographically but they are more
isolated than the large, continental co-clusters.

\subsection{English n-grams}


Our GTSC framework clusters the $3$-gram and $4$-gram tensors into $486$ and
$383$ groups, respectively.  We rank the groups by decreasing order of the
popularity score described in Section~\ref{sec_rank_clusters}.  We find several
conclusions that hold for both tensor datasets.  Highly ranked groups are mostly
composed of stop words (i.e., common words) such as \textit{the, a, we, is, by} (we define
stop words as those most connected with other words).
In fact, the top two groups consist nearly entirely of stop words (see Table~\ref{table_stop}).  In addition, 48\% ($3$-gram) 
and 64\% ($4$-gram) of words in the first group are prepositions (e.g., \text{in, of, as, to})
and link verbs (e.g., \textit{is, get, does}).  In the second group, 64\% ($3$-gram) and 57\%
($4$-gram) of the words are pronouns (e.g., \textit{we, you, them}) and link verbs.  This
result matches the structure of English language where link verbs can connect
both prepositions and pronouns whereas prepositions and pronouns are unlikely to
appear in close vicinity.  Since prepositions are more common than pronouns,
they rank first.

\begin{table}[tbp]
\caption{Fraction of words in the top two groups (in terms of popularity, see Section~\ref{sec_rank_clusters}) that are among the top 100 (200) most frequently used words in
the English (Chinese) written language corpora.}
\vspace{-0.5cm}
\label{table_stop}
\begin{center}
\begin{tabularx}{\linewidth}{>{\hsize=1.1\hsize}X >{\hsize=0.8\hsize}Z >{\hsize=0.8\hsize}Z >{\hsize=0.8\hsize}Z >{\hsize=0.8\hsize}Z}
\toprule
 &\multicolumn{2} {>{\hsize=2\hsize}Z}{English} &\multicolumn{2} {>{\hsize=2\hsize}Z}{Chinese} \\
\midrule
Groups & $3$-gram & $4$-gram & $3$-gram & $4$-gram \\
\midrule
1st Group & $24/29$ & $11/11$ &$42/84$ & $31/31$\\
2nd Group  & $23/25$ & $42/47$ & $33/74$ & $23/36$\\
\bottomrule
\end{tabularx}
\end{center}
\end{table}

%

Groups ranked in the middle mostly consist of semantically related English
words.  For instance, a few of these groups are \{\textit{cheese, cream, sour,
  low-fat, frosting, nonfat, fat-free}\}, \{\textit{bag, plastic, garbage,
  grocery, trash, freezer}\}, \{\textit{infection, chronic, sexually,
  transmitted, diseases, hbv}\}.  The lowest ranked groups are comprised mostly
of non-English words that appear in English text.  For example, one such group is \{\textit{je, ne, sais, quoi}\}, which is a French
phrase.  We may consider these groups as outliers.




The clustering the $4$-gram tensor contains some groups that the $3$-gram
tensor fails to find.  For example, one cluster is \{\textit{german, chancellor,
  angela, merkel, gerhard, schroeder, helmut, kohl}\}.  Angela Merkel, Gerhard
Schroeder, and Helmut Kohl have all been German chancellors, but it requires a
$4$-gram to make this connection strong.  Likewise, some clusters only appear
from clustering the $3$-gram tensor.  One such cluster is \{\textit{church,
  bishop, catholic, priest, greek, orthodox, methodist, roman, priests,
  episcopal, churches, bishops}\}.  In $3$-grams, we may see phrases such as
``catholic church bishop", but $4$-grams containing these words likely also
contain stop words, \emph{e.g.}, ``bishop of the church".  However, since stop
words are in a highly ranked cluster, the connection is destroyed once the top
clusters are formed.

\subsection{Chinese n-grams}

We find that many of the conclusions from the English $n$-gram datasets also
hold for the Chinese $n$-gram datasets. For example, many of the words in the
top two groups are stop words (see Table~\ref{table_stop}).  Groups scoring in
the middle consist of semantically similar words, and low-scoring groups
are foreign language phrases that appear in Chinese literature.


However, there are also several differences from the English $n$-gram data.
First, we find that groups in Chinese are usually larger than those in English.
The average sizes of small-to-medium groups (3--100 words) is 6.8 for
English and 10 for Chinese.  We suspect this pattern is due to the fact that
Chinese words are more semantically dense than English words.  For example, it
is common that Chinese words have multiple meanings and can co-occur with a
large number of words.


We note that there are several words form the top two groups that are
not typically considered as stop words.  For example, \textit{society, economy,
  develop, -ism, nation, government} are among the top 100 most common words.
(These are translations of the Chinese words).  This is a consequence of the
dataset coming from scanned Chinese-language books and is a known issue with
the Google Books corpus~\cite{pechenick2015characterizing}.  In this case, it is a feature as we are 
illustrating the efficacy of our tensor clustering framework rather than making
any linguistic claims.

\subsection{Enron Email Tensor}
\begin{figure}[tb]
\begin{center}
  \includegraphics[width=\linewidth]{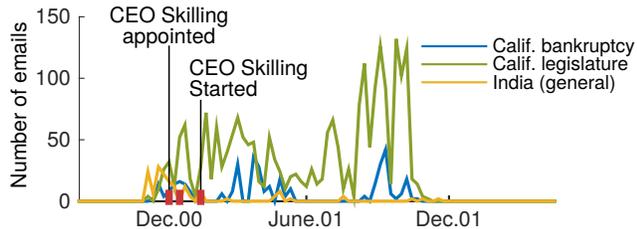}
\vspace{-0.15cm}  
  \caption{Enron email volume on three labeled topics.
  Our GTSC framework finds a co-cluster consisting of these three topics at
  the time points labeled in red, which seems to correlate with various
  events involving the CEO.}
  \label{fig_3topic}
\end{center}
\end{figure}


In total, the algorithm finds $23$ co-clusters of topics, people, and time.  The
most popular group corresponds to three topics, 19
people, and 0 time intervals.  Similar to the $n$-grams and airport-airline
data, this cluster corresponds to high-volume entities, in this case common
topics and people who send a lot of emails.  The three topics are ``Daily
business", ``too few words", and ``no matching topic", which account for roughly
90\% of the total email volume.  (The latter two topics are capturing outliers
and emails that do not fall under an obvious category).  The 19 employees
include 11 managers:~the CEO, (vice) preseidents, and directors.  These
employees are involved in 42\% of the total emails.  There is no time interval
in this co-cluster because these high-volume topics and employees are balanced
throughout time.

%
%

We also found several interesting co-clusters. One
consists of the topics ``California bankruptcy", ``California
legislature", and ``India (general)", during three weeks in December 2000 and January
2001, and 13 employees.  These time points correspond to various events
involving CEO Skilling (Figure~\ref{fig_3topic}).  Each of the 13 employees in the
co-cluster sent at least one email from at least one of the topics.  Another
co-cluster consists of the topics ``General newsfeed", ``Downfall newsfeed", and
``Federal Energy Regulatory Commission/Department of Energy" and several weeks
from March 2001 and December 2001.  These time intervals coincide with investor
James Chanos finding problems with Enron in early 2001 and the serious financial
troubles encountered by the company in late 2001.  


\section{RELATED WORK}

There are a variety of methods for tensor decomposition that could be used to cluster a tensor~\cite{cao2013robust,huang2008simultaneous}.  For instance, Huang
et al.\ use a higher-order SVD as a basis for
clustering~\cite{huang2008simultaneous} and Anandkumar et al.~\cite{Anandkumar-2014-tensor-decomp} use a latent-variable model.  We studied standard algorithms for the PARAFAC decomposition in this light and found that our tensor clustering method outperformed this general strategy. 

Our work provides a technique for ``co-clustering", i.e, finding clusters of
different entities through their interactions.  Prior work in this area
has focused on finding groups of indices in the rows and columns of a matrix of
interaction data~\cite{dhillon2001co,dhillon2003information}.  Our work provides
the first generalization of these techniques for the higher-order interactions
encountered in tensor data.  We note that recent work has looked at
co-clustering several entity types
simultaneously~\cite{gao2005consistent,bao2013social}.  However, these methods
are still based on first-order information (graph structure).


Our GTSC framework is immediately applicable as a clustering method for
multiplex networks~\cite{newman2003structure}.  Indeed, we found interesting
structure in the airline-airport dataset, which is an example of a multiplex
network.  Related work in this area includes multi-view
clustering~\cite{zhou2007spectral}, ensemble methods~\cite{fern2004solving}, and
multi-network clustering~\cite{ni2015flexible}.  However, these methods are
specialized for multiplex networks, whereas our method provides a general tensor
clustering framework.


\section{CONCLUSION}

Tensors are increasingly common in modern applications and clustering tensor
data is fundamental for discovering patterns and data analysis. However, tensor
clustering is a difficult task for two reasons:~higher-order structure in tensors
is difficult to model and obvious extensions of models to higher-order data are
computationally challenging.  In this paper we proposed the General Tensor
Spectral Co-clustering (GTSC) method. Our method
addresses these issues by modeling higher-order data with a new stochastic
process, the super-spacey random walk, which is a variant of a higher-order
Markov chain. Our iterative solver and sweep cut procedure for biased
conductance can achieve near-linear complexity in the number of non-zeros in
tensors. In synthetic experiments our GTSC out-performs state-of-the-art
spectral methods and tensor decomposition methods and can efficiently handle
skew in the distribution of the indices of the non-zeros in the tensor.
Furthermore, our GTSC framework can find clear cluster structures in various
tensor datasets, including English and Chinese $n$-gram text, an airline and
airport multiplex network, and Enron email data.

In terms of future work, we'd like to create tensors that bridge information from multiple modes. For instance, the clusters from the 3-gram data were different from the 4-gram data and it would be useful to have a holistic tensor to jointly parition both 3- and 4-gram information. This is important because some of the clusters in the $n$-gram data correspond to automatically extracted knowledge, such as the cluster with the names of various German chancellors. This aspect of our output also merits further investigation as it would require overlapping clusters to be useful to knowledge extraction efforts.

\fontsize{8}{9}\selectfont

\textbf{Acknowledgements.} This work was supported by NSF IIS-1422918 and DARPA SIMPLEX. 
ARB is supported by a Stanford Graduate Fellowship.
We are grateful to our colleagues who helped revise early drafts.

\bibliographystyle{abbrv}
\bibliography{refs}


\appendix


\section{STATIONARY DISTRIBUTION}
\label{app_stationary}
We provide a heuristic proof here and a reference for how to make it formal. Suppose the process has run for a very long time and that $\vx_t$ is the current empirical distribution. From equation~\eqref{equ_spsurfer}, we have 
\[ 
 \text{Pr}(X_{t+1}=i | X_{t} = j) = (1-\alpha) \vv_i + \alpha \bigl( \sum_{\mathclap{(j,k) \in \mathcal{F}}} \elm{P}_{i,j,k} \vx_t(k) + \sum_{\mathclap{(j,k) \not\in \mathcal{F}}} \vx_t(k) \vx_t(i) \bigr).
\]
Because the process has run for a very long time, the empirical distribution $\vx_t$ is essentially fixed. 
Thus, the stochastic proecess can be approximated by a Markov chain with the transition matrix:
\[ 
 \alpha \cmP[\vx_t] + \alpha \vx_t( \ve^T - \ve^T \cmP[\vx_t] ) + (1-\alpha) \vv \ve^T.
\]
At stationarity of the super-spacey random walk, the stationary distribution of this Markov chain must be equal to $\vx_t$. A more formal proof is to use the results from \cite{Benaim-1997-vrrw} to show that (\ref{equ_spmp}) is the necessary condition of the stationary distribution in a vertex reinforced random walk.

\section{PROOF OF THEOREM \ref{THEO_UNIQ}}
\label{app_uniq}

Let $\mR$ denote the mode-$1$ unfolding of $\cmP$:
\begin{equation*}
\mR = [\cmP(:,:,1) \mid \cmP(:,:,2) \mid \cdots \mid \cmP(:,:,n)].
\end{equation*}
Note that $\mR (\vx \kron \vx) = \cmP \vx^2$ where $\kron$ is the Kronecker product. 
Assume $\vx$ and $\vy$ are two solutions of \eqref{equ_spmp}. Let $r_x = \normof[1]{\mR (\vx \kron \vx)}$ and $r_y = \normof[1]{\mR (\vy \kron \vy)}$. Then 
\[ \normof[1]{\vx - \vy} \le \alpha \normof[1]{\mR (\vx \kron \vx - \vy \kron \vy)} + \alpha \normof[1]{ (1-r_x) \vx + (1-r_y) \vy }. \]
By Lemma 4.5 of \cite{gleich2015multilinear}, the first term
\[ \alpha \normof[1]{\mR (\vx \kron \vx - \vy \kron \vy)} \le \alpha \normof[1]{\mR} \normof[1]{\vx \kron \vx - \vy \kron \vy} \le 2 \alpha \normof[1]{\vx - \vy}. \]
The second term satifies 
\[ \begin{aligned}
 & \alpha \normof[1]{( 1 - r_x)(\vx - \vy)} + \alpha |r_y - r_x| \\
 & \qquad \le \alpha \normof[1]{\vx -\vy} + \alpha \normof[1] {\mR (\vx \kron \vx - \vy \kron \vy) } \le 3 \alpha \normof[1]{\vx - \vy}.
\end{aligned} 
\]
Combining the above two facts, we know when $\alpha<1/5$ the solution is unique. For an $m$-mode tensor, this idea generalizes to $\alpha < 1/(2m-1)$. 

For the convergence of the fixed point algorithm \eqref{eq:iter}, the same analysis shows that $\normof[1]{\vx_{k+1} - \vx^*} \le 5\alpha \normof[1]{\vx_{k} - \vx^*}$, and so the iteration converges when the solution is unique.

\end{document}